\documentclass[%
 reprint,
 amsmath,amssymb,
 aps,
]{revtex4-1}

\usepackage{graphicx}
\usepackage{dcolumn}
\usepackage{bm}

\begin{document}

\preprint{APS/123-QED}

\title{Tailoring a nanofiber for enhanced photon emission and coupling efficiency from single quantum emitters}

\author{Wenfang Li}
\author{Jinjin Du}%
 \email{Email:jinjin.du@oist.jp}
 \author{S\'ile Nic Chormaic}
\affiliation{%
 Light-Matter Interactions Unit, Okinawa Institute of Science and Technology Graduate University, Onna, Okinawa 904-0495, Japan
}%

\date{\today}

\begin{abstract}
We present a novel approach to enhance the spontaneous emission rate of single quantum emitters in an optical nanofiber-based cavity by introducing a narrow air-filled groove into the cavity. Our results show that the Purcell factor for single quantum emitters located inside the groove of the nanofiber-based cavity can be at least six times greater than that for such an emitter on the fiber surface when using an optimized cavity mode and groove width. Moreover, the coupling efficiency of single quantum emitters into the guided mode of this nanofiber-based cavity can reach up to $\sim$ 80 $\%$ with only 35 cavity-grating periods. This new system has the potential to act as an all-fiber platform to realize efficient coupling of photons from single emitters into an optical fiber for quantum information applications.  

\end{abstract}

\pacs{Valid PACS appear here}
\maketitle

Micro- and nanoscale optical systems are receiving increasing attention from the quantum optics community. Owing to their capacity to create an efficient, scalable interface between light and matter at the single photon and single emitter levels, microscale systems are of fundamental importance for quantum information science \cite{1,2,3,4}. These systems should have an ultra-small mode volume to increase the spontaneous emission rate of single quantum emitters and should be able to couple fluorescent photons to a single mode fiber with high efficiency. In recent years, optical nanofibers (ONFs) with sub-wavelength diameters have attracted great interest from researchers as useful light-matter interface platforms \cite{5,6} benefiting from the tight confinement of the evanescent field around the fiber, the unique geometries provided by fiber modes, and low optical loss \cite{7,8}. They allow for the efficient coupling of fluorescent photons emitted by single emitters near the ONF surface to a propagating mode of the nanofiber and subsequently to the guided mode of a standard single-mode fiber   through an adiabatically tapered region. Efficient coupling of photons from a variety of quantum emitters, such as laser-cooled atoms, has been experimentally demonstrated \cite{9,10,11,12,13,14}. As theoretically expected, about 8$\%$ of the guided light can be absorbed when an atom is trapped about 200 nm away from the nanofiber surface and this decreases gradually with increasing  atom-surface distance \cite{11,15,16}. For solid-state-based quantum emitters \cite{17}, such as semiconductor quantum dots \cite{18,19}, nitrogen-vacancy (NV) centers in diamond \cite{20,21,22} and even emitters in 2D materials \cite{23}, experimental progress has demonstrated efficient positioning of such single quantum emitters onto the surface of a nanofiber. For single emitters located on a nanofiber surface,  $\sim$ 20$\%$ of the fluorescent photons can be coupled into the fiber's guided modes \cite{19}. Various proposals for solid-state-based emitters based on optical nanofibers have been made with the aim of improving the coupling efficiency \cite{24,25,26,27,28,29}, especially ideas using nanofiber-based cavities \cite{30,31}. With the aid of such cavities, the coupling efficiency and the spontaneous emission rate of single-quantum emitters can be further increased because of the enhanced interaction of the emitter with its environment (Purcell effect) \cite{32}. It has been predicted that the coupling efficiency could be close to unity by using nanofiber-based cavities \cite{30}. Therefore, nanofiber-based cavity systems combined with ONFs and fiber in-line cavities have been proposed and fabricated utilizing various technologies \cite{33,34,35,36,37,38}. In addition, an enhancement of the spontaneous emission rate of single quantum dots based on such a system has been experimentally demonstrated \cite{39,40}. The aforementioned systems are for emitters located on the surfaces of nanofibers; however, further improvement of the interactions between emitters and guided modes are mostly modest for silica nanofibers. Thus, it is still challenging to achieve a significant enhancement of the fluorescence from single emitters coupled to fiber-based cavities with a high quality factor, Q , and a small effective mode volume, $V_{eff}$ , as well as a high coupling efficiency.

In this Letter, we propose a tailored optical nanofiber (TONF) for enhanced photon emission and coupling efficiency from single-quantum emitters. This tailored nanofiber is a nanofiber-based cavity \cite{41} with a narrow, air-filled groove inside the cavity (Fig. 1). In contrast to a simple slotted ONF for atom trapping \cite{42}, introducing this groove not only allows for single emitters to reside in the strongest intensity field of the nanofiber, but it also induces an increase in the normalized maximum electric field, which significantly enhances the spontaneous emission rate of single emitters inside the cavity. For the TONF with a moderate groove width of 50 nm, the coupling efficiency for single emitters can reach 80$\%$ with a short-period number of air-nanohole arrays on the nanofiber and the Purcell factor can be as high as $\sim$90 . The dielectric discontinuities inside the cavity induced by this narrow air-filled groove can result in a 10$\times$ increase in the Purcell factor compared with the value on the cavity surface. These results pave the way for future experiments exploiting the Purcell-enhanced interaction between single emitters and the optical nanofiber-based cavity with a high coupling efficiency.

\begin{figure}[htb]
\centerline{
\includegraphics[width=8.5cm]{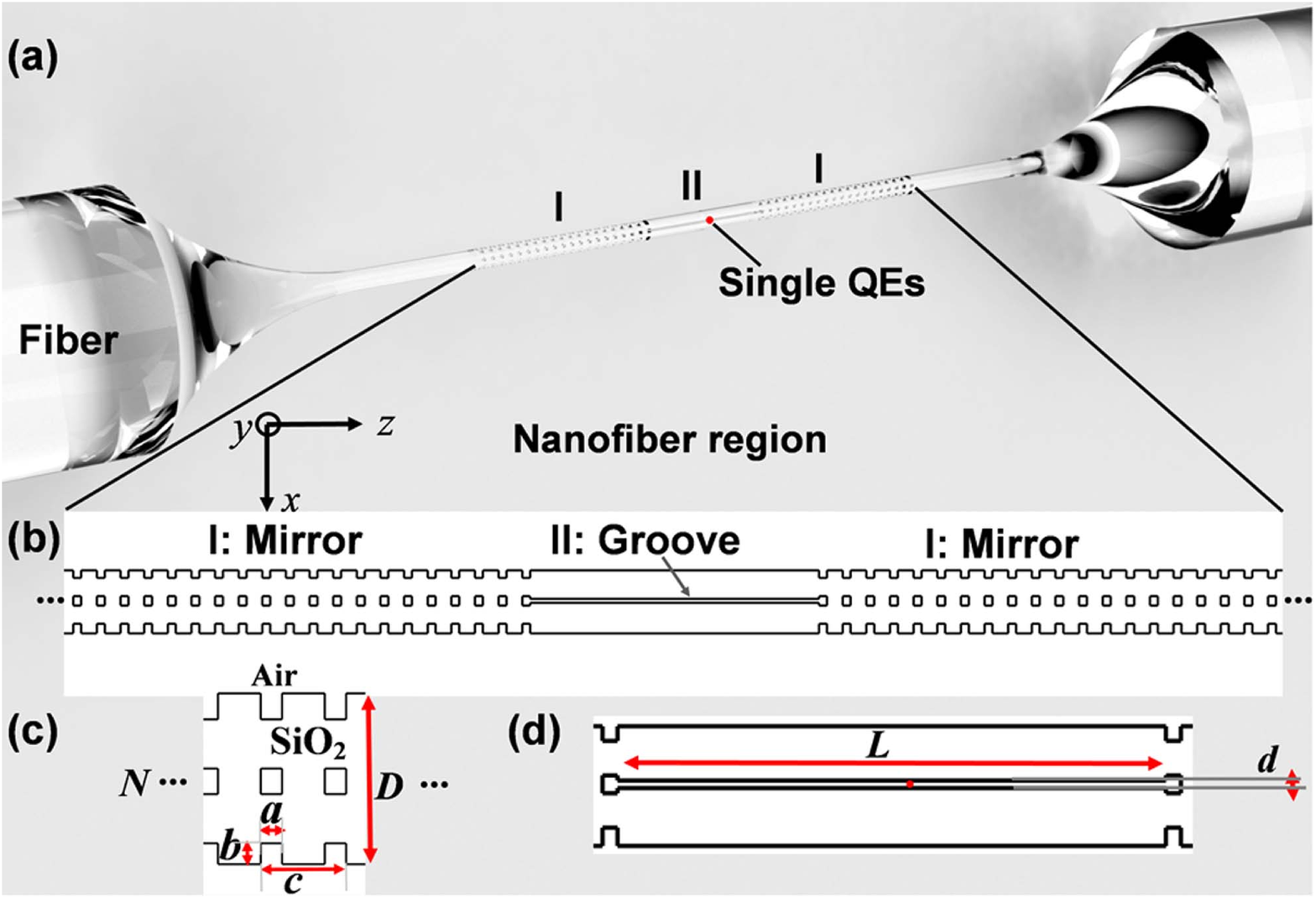}}
\caption{ (Color online) The proposed system for a tailored-nanofiber based cavity (TONF) coupled to a single quantum emitter. (a) Schematic of the proposed coupled system, mainly composed of single quantum emitters (QEs), cavity mirrors I and a narrow groove II inside the cavity region in the waist of a nanofiber, with geometry shown in (b)-(d). (b)Enlarged view of the main structures. (c) The geometry of structure I with an etch width $a=100$nm, an etch depth $b=100$nm, a period length $c=310$nm, and an etch number $N$ as well as fiber diameter $D$. (d) The geometry of structure II with a groove length of $L$=2.2$\mu m$ and width of $d$. }
\label{fig:false-color}
\end{figure}

The TONF system shown in Fig.1 consists of two main parts: an optical nanofiber-based cavity induced by periodic air-nanohole arrays (structure I) and an air-filled groove embedded in the cavity region (structure II). Hence, the optical characteristics of this proposed structure are determined by both structures. We previously fabricated an optical nanofiber-based cavity induced by periodic air-nanohole arrays \cite{41} with a high Q factor with very few grating periods. We used the same parameters for structure I as in our previous work (Fig. 1). When a narrow air-filled groove is introduced into the cavity region, the electric field distribution along the nanofiber for a polarized input mode is displayed in Fig. 2(a), obtained using a finite element method (FEM) based on COMSOL Multiphysics. An enlarged region of the groove is shown in Fig. 2(b) for two different cases, i.e. single emitters on the surface and in the groove. It shows the cross-sections of the normalized electric field distribution along the fiber (axial direction) for y- polarized modes with a nanofiber diameter $D$  of 800 nm and a groove width of 50 nm. The optical response spectrum, such as the transmission through the TONF, has been calculated by using a commercial finite-difference time-domain (FDTD) method performed on a commercial package (FDTD Solutions, Lumerical); see Fig. 2(c). Based on the resonant peaks for $N$=35, the Q factors of the x- and y-polarized modes are $\sim$1578 and $\sim$796, respectively.

\begin{figure}[htb]
\centerline{
\includegraphics[width=\linewidth]{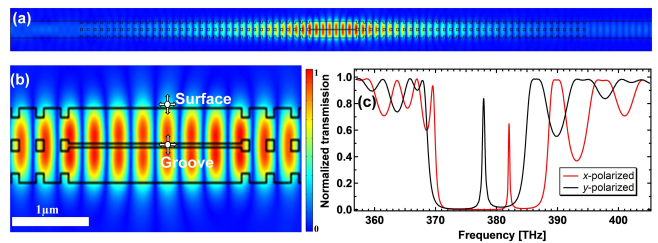}}
\caption{ (Color online) Optical responses of a nanofiber-based cavity with a narrow air-filled groove (TONF) with dimensions ($D$, $N$, $d$)=(800, 35, 50) nm. (a) Electric field distribution along the nanofiber for y-input polarization. (b) The field distribution near the groove of the TONF. Two locations of single emitters are considered: on the surface and in the groove. (c) The normalized transmission of the TONF for x- (red curve) and y-(black curve) polarized input mode.}
\label{fig:false-color}
\end{figure}

The air-filled groove embedded in the TONF induces local field enhancement in the normalized maximum field using dielectric discontinuities with subwavelength scales\cite{43}. From Maxwell’s equations, normal components of the electric displacement are continuous across the boundary of two different dielectrics, that is ${\varepsilon _{silica}}{E_{silica}} = {\varepsilon _{air}}{E_{air}}$\cite{44}. The intensity of the electric field distribution $\left| {\rm{E}} \right|$  of the quasi x- and y-polarized modes in the TONF for different groove widths, $d$, is shown in Fig. 3. The peak of the electric field $\left| {\rm{E}} \right|$  for x- polarized modes at the cavity is enhanced ~2x compared to the case without a groove inside the cavity (Fig. 3(a)). From Fig. 3 (b), for y- polarized modes, there is no significant distinction between the electric field distribution in the cavity with and without a groove. Compared to the value of the electric field on a fiber surface, the electric field in the groove increases many fold for both the x- and y-polarized modes.

\begin{figure}[htb]
\centerline{
\includegraphics[width=9cm]{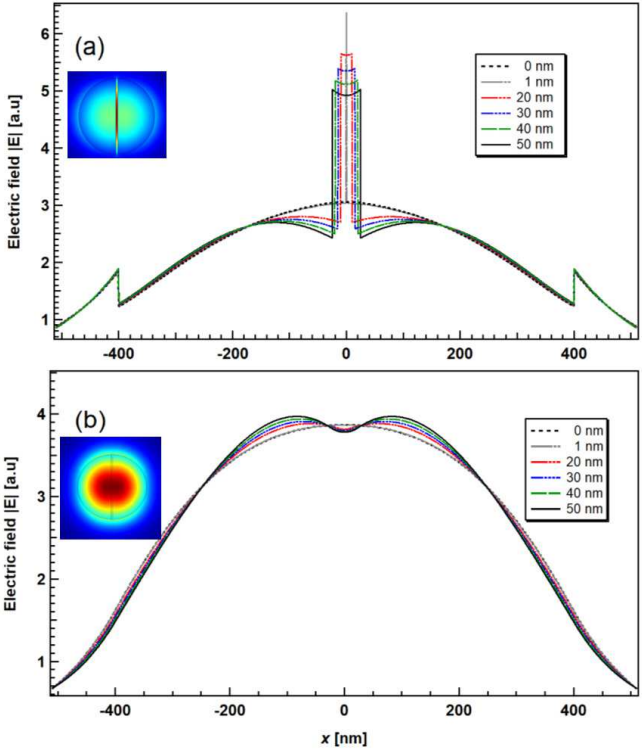}}
\caption{(Color online) The intensity of the electric field distribution of the quasi (a) x- and (b) y-polarized modes, in the TONF for different groove widths, $d$. Compared to the case of no groove (black -dotted curve), fields with widths of 20 nm (red-dashed curve), 30 nm (blue-dashed curve), 40 nm (green-dashed curve), and 50 nm (black curve) are displayed. The optimal enhancement of electric filed by an air-filled groove is achieved using a setting $d$ of 1 nm.}
\label{fig:false-color}
\end{figure}

The spontaneous decay rate of a quantum emitter can be enhanced when it is coupled to a TONF based on the Purcell effect (see Figure 1(a)), which was first described by E. M. Purcell in 1946 \cite{32}. In a TONF, i.e., a nanofiber-based cavity for which the resonant wavelength of the quantum emitter $\lambda $ is tuned to the wavelength of the guided mode, the Purcell factor, F, is given in terms of the cavity quality factor, Q, and the effective mode volume $V_{eff}$, such that $F = 6Q{(\lambda /2n)^3}/({\pi ^2}{V_{eff}})$ for an emitter located at the peak of the electric field with the index of refraction $n$. Based on Fermi’s golden rule, modification of the spontaneous decay rates of single emitters is proportional to the number of local density of states (LDOS) of available final photon states. The  LDOS ${\rho _p}$ for the electromagnetic modes is directly proportional to the imaginary part of the corresponding electromagnetic Green’s function. The spontaneous decay rate $\gamma $ of a quantum emitter in the presence of the TONF can be given by \cite{45},
\begin{equation}
\begin{array}{l}
\gamma  = \frac{{\pi {\omega _0}}}{{3\hbar {\varepsilon _0}}}{\left| {\bf{p}} \right|^2}{\rho _p}({{\bf{r}}_0},{\omega _0}),\\
{\kern 1pt} {\rho _p}({{\bf{r}}_0},{\omega _0}) = \frac{{6{\omega _0}}}{{\pi {c^2}}}\left[ {{{\bf{n}}_p} \cdot {\mathop{\rm Im}\nolimits} \left\{ {{\bf{\mathord{\buildrel{\lower3pt\hbox{$\scriptscriptstyle\leftrightarrow$}} 
\over G} }}({{\bf{r}}_0},{{\bf{r}}_0};{\omega _0})} \right\} \cdot {{\bf{n}}_p}} \right]
\end{array}
\label{eq:refname1}
\end{equation}
where ${\omega _0}$ is the transition angular frequency, ${\left| {\bf{p}} \right|^2}$ is the transition dipole moment, ${\bf{p}} = \left| {\bf{p}} \right|{{\bf{n}}_p}$ is the dipole moment of an emitter with ${{\bf{n}}_p}$being the unit vector in the direction of $p$, and ${{\rm{\mathord{\buildrel{\lower3pt\hbox{$\scriptscriptstyle\leftrightarrow$}} 
\over G} }}({{\bf{r}}_0},{{\bf{r}}_0};{\omega _0})}$ is the dyadic Green’s function calculated at the emitter’s location ${{\bf{r}}_0}$. To characterize the coupling between guided modes and a quantum emitter placed in the groove, the coupling efficiency  $\beta  = {\gamma _{guided}}/{\gamma _{total}}$ is defined as the ratio between the emission from single emitters into the guided mode of a TONF ${\gamma _{guided}}$  and the total decay rate ${\gamma _{total}}$ into all radiative channels in the presence of the TONF. Moreover, the emission enhancement factor can be written as $\zeta  = {\gamma _{guided}}/{\gamma _0}$ with ${\gamma _0}$ being the intrinsic spontaneous emission rate into free space. Therefore the total enhancement of the spontaneous emission from a single quantum emitter, namely, the Purcell factor, can be deduced as $F = \zeta /\beta $. Using the 3D finite-different time-domain (FDTD) method, the spontaneous decay rate $\gamma $ and coupling efficiency ${\beta }$ can be obtained by calculating the Green’s tensor and the power emission into cavity modes.

\begin{figure}[htb]
\centerline{
\includegraphics[width=9cm]{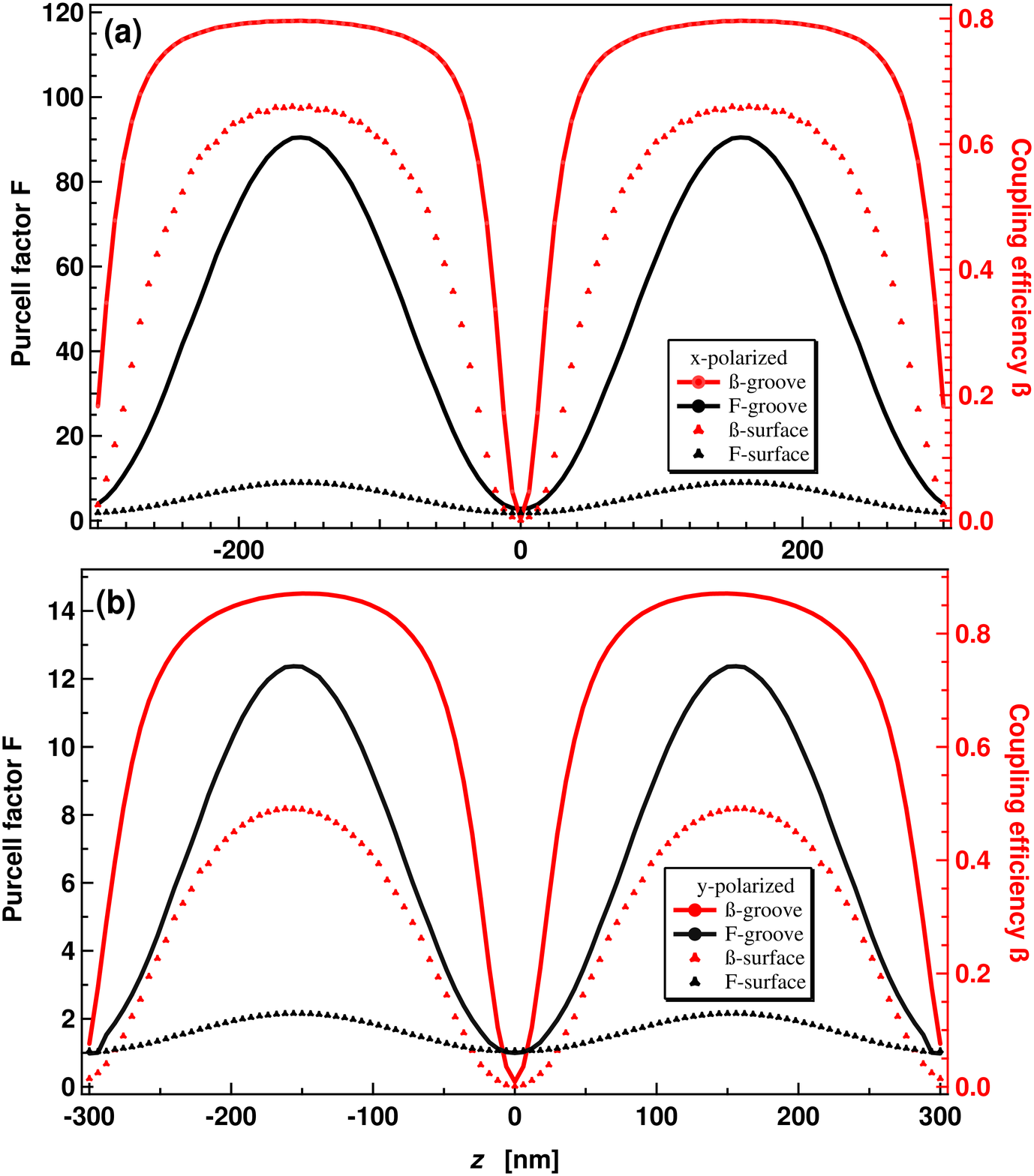}}
\caption{(Color online) Coupling efficiency ${\beta }$ and the Purcell factor F for different emitter locations (see Fig.2 (a) (b)) of the TONF along extending direction of the groove at $d$= 50 nm. For an emitter in the middle of the groove, solid red lines indicates coupling efficiency ${\beta }$ and solid black lines are for the Purcell factor F. Accordingly red- and black triangles are for an emitter on the surface of nanofiber.}
\label{fig:false-color}
\end{figure}

\begin{figure}[htb]
\centerline{
\includegraphics[width=\linewidth]{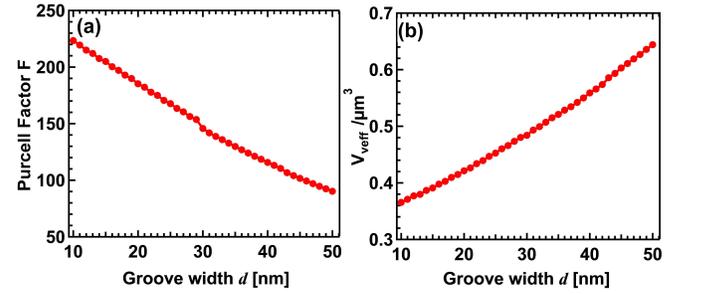}}
\caption{(Color online) Dependence of Purcell factor F and effective mode volume $V_{eff}$ on the groove width $d$. (a) indicates the Purcell factor F (red-cycle and line) and (b) is the effective mode volume (red-cycle and line).}
\label{fig:false-color}
\end{figure}

 Figure 4 characterizes the coupling efficiency ${\beta }$ (solid red line) from single QEs and the corresponding Purcell factor F (solid black line) for an emitter positioned at the center of the groove along the z-axis. For comparison, the red- and black triangles in Fig .4(a) (b) represent the corresponding cases for an emitter on the surface of the nanofiber. With a groove of width $d$=50 nm and a period number $N$ of 35, the Purcell factor F can reach 90 for x- polarized modes when an emitter is positioned at the peak intensity of the field in the groove; this is several times larger than that of the y- polarized modes owing to the field enhancement induced by the narrow air-filled groove. Compared with an emitter on the surface of the TONF, the Purcell factor F for both of the polarized modes increases by at least 6$\times$. The coupling efficiency ${\beta }$ of both polarized modes with the groove can reach $\sim$80$\%$, which is larger than that for the emitter on the surface of the TONF.

 \begin{figure}[htb]
\centerline{
\includegraphics[width=\linewidth]{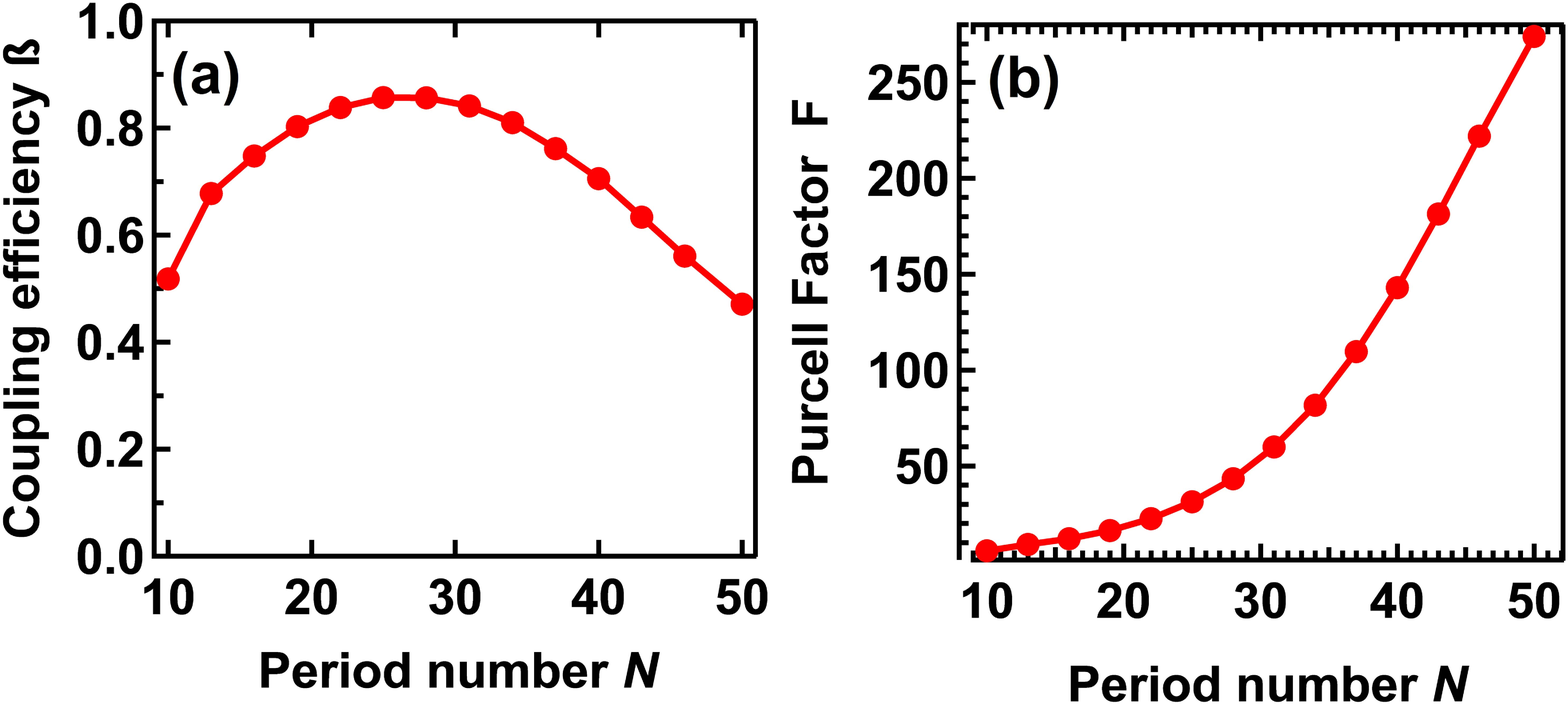}}
\caption{(Color online) Dependence of coupling efficiency ${\beta }$ and Purcell factor F on the period number $N$. (a) the change of coupling efficiency ${\beta }$ (red-cycle and line) and (b) is for the Purcell factor F (red-cycle and line).}
\label{fig:false-color}
\end{figure}
   
In principle, the air-filled groove width $d$ should be as small as possible to achieve the highest possible local field enhancement for x- polarized modes. Fig. 5 shows the change in the Purcell factor F and the effective mode volume $V_{eff}$ with changes in the groove width $d$, respectively. As expected, F gradually increases with decreasing width $d$. In particular, F can reach as much as 220 with $d$ = 10 nm. To get as small an effective volume $V_{eff}$  as possible, the width $d$ should be as small as possible. It shows that the effective mode volume can be further reduced by half when the width $d$ diminishes to zero. However, considering the practical size of an emitter such as a quantum dot, with a diameter of $\sim$6 nm, and the technical limitations on positioning it within the groove during an experiment, we choose $d$ =50 nm for the aforementioned calculation.

For further optimization of ${\beta }$ and F, we calculated their dependence on the period number $N$ ( Fig. 6). F increases with the increasing period number $N$ and F can reach 250 at $N$ =50. Additionally, the coupling efficiency ${\beta }$ presents a trend from rising to declining with increased period numbers $N$ due to the increased grating loss \cite{31} which could be reduced by designing a modulated Bragg mirror \cite{46}. Consequently, there is a tradeoff between the possible enhanced spontaneous decay rate of a quantum emitter denoted by F, and the possibility of coupling fluorescence photons from single emitters to the TONF, measured by the coupling efficiency ${\beta }$. The efficiency ${\beta }$ reaches a maximum value for approximately 30 periods.

In summary, we have explored a tailored optical nanofiber (TONF) as a nanoscale cavity QED system that could yield a high photon collection efficiency of ${\beta }$$\sim$ 80$\%$  and an enhancement of the spontaneous radiation decay rate for a single quantum emitter of F$\sim$ 90 using a moderate groove width of 50 nm. We have shown that the TONF with the narrow air-filled groove is a useful system for efficient coupling of fluorescence photons from single emitters into an optical fiber and should provide enhanced interaction between single quantum emitters and the guided mode of a nanofiber. This system could be used for observing strong light-matter interactions and multicolor quantum manipulation such as atom trapping for quantum information processing. 
 
 \mbox

This work was supported by Okinawa Institute of Science and Technology Graduate University.The authors would like to thank F. Le Kien for invaluable discussions. S.Aird are acknowledged for comments on the manuscript.

\end{document}